Article

# A unified solution to the low-mass X-ray binary diversity puzzle: saturated magnetic braking coupled with irradiation-driven winds

Zhu-Ling Deng [a], Xiang-Dong Li [b,c,*], Bo Wang [a,*], Philipp Podsiadlowski [d,e], Yong Shao [b,c], Dong-Dong Liu [a], Zhan-Wen Han [a]

[a] International Centre of Supernovae (ICESUN), Yunnan Key Laboratory of Supernova Research, Yunnan Observatories, Chinese Academy of Sciences, Kunming 650216, China
[b] School of Astronomy and Space Science, Nanjing University, Nanjing 210023, China
[c] Key Laboratory of Modern Astronomy and Astrophysics, Nanjing University, Ministry of Education, Nanjing 210023, China
[d] London Centre for Stellar Astrophysics, London SW8, UK
[e] University of Oxford, St Edmund Hall, Oxford OX1 4AR, UK

ARTICLE INFO

*Article history:*
Received 20 April 2026
Received in revised form 24 June 2026
Accepted 29 July 2026
Available online 28 August 2026

ABSTRACT

The standard magnetic braking (MB) model has long failed to explain key properties of low-mass X-ray binaries (LMXBs) — including their accretion rates, the formation of ultracompact binaries, and orbital expansion — casting doubt on binary evolution and gravitational-wave progenitor predictions. Phenomenological modifications exist, but a unified, physically grounded mechanism has been missing. Here we show that the overlooked coupling between magnetic saturation and irradiation-driven winds (IDW) provides this missing framework. Accretion-powered irradiation enhances the donor wind, which in turn amplifies MB exactly where needed. Our Saturated MB + IDW (SMB + IDW) model simultaneously resolves three outstanding discrepancies with no fine-tuning: (1) it reproduces the full observed distribution of accretion rates across persistent and transient LMXBs; (2) it naturally explains ultracompact X-ray binary formation; and (3) it quantitatively matches the orbital period distribution of descendant millisecond pulsar (MSP) binaries — a critical independent test failed by all previous models. Critically, the SMB + IDW model transforms MB from an externally imposed condition into a self-consistently regulated process tied to the accretion history. By unifying LMXB evolution to MSPs in a single physical picture, our work resolves the long-standing diversity puzzle and establishes a physics-based foundation for next-generation population synthesis, gravitational-wave progenitor predictions, and compact binary interpretation across cosmic time.

## 1. Introduction

Magnetic braking (MB) extracts stellar angular momentum (AM) through the coupling of the stellar magnetic field with the stellar wind [1–3] and drives the observed spin-down of Sun-like stars [4]. In close binaries, tidal synchronization forces stellar rotation to match the orbital motion, allowing MB to directly remove orbital AM [5,6]. Consequently, for low-mass X-ray binaries (LMXBs), MB governs the long-term orbital evolution through AM loss and, by setting the mass-accretion rate, determines the system's X-ray luminosity [7–9], thereby positioning MB as the central mechanism.

The standard MB prescription applied to LMXBs, which is rooted in the spin-down of solar-type stars [4,10] and formalized by Verbunt et al. [5] and Rappaport et al. [6], is inconsistent with key observations. It predicts accretion rates for persistent LMXBs that are 1–2 orders of magnitude lower than observed values [7,9,11]. Reproducing the observed populations of tight-orbit millisecond pulsar (MSP) helium white dwarf (WD) binaries and ultracompact X-ray binaries (UCXBs) requires significant fine-tuning of initial parameters [8,11–13]. Furthermore, it fails to explain the anomalous orbital expansion observed in a subpopulation of LMXBs [14,15].

Several modified MB prescriptions have been developed to address these discrepancies. Models such as the $\tau$-boost [9] and CARB [16] successfully match the accretion rates of persistent LMXBs, alleviate the need for fine-tuning [11,17], and account for anomalous orbital expansion in certain systems [15]. A common limitation, however, is their difficulty in simultaneously explaining the low accretion rates of transient sources and the observed population of wide-orbit MSP binaries within a single framework [11,17,18]. Thus, while these studies have broadened the phenomenological scope of MB models, a unified theory that explains the full range of LMXB properties — from persistent and transient sources to compact and wide systems — remains elusive.

* Corresponding authors.
*E-mail addresses:* lixd@nju.edu.cn (X.-D. Li), wangbo@ynao.ac.cn (B. Wang).

A potential solution emerges from two converging lines of evidence. First, stellar observations indicate that magnetic activity saturates above a critical rotation rate [19–21], leading to the concept of saturated MB, where the torque depends only weakly on rotation [22–24]. This concept is supported by the orbital period distribution of detached low-mass binaries [25,26] and recent studies of contact binaries [27].

Second, irradiation profoundly affects companion stars, creating a substantial temperature contrast between hemispheres [28,29] — exemplified by systems such as AA Dor (∼20,000/ ∼2000 K; Hilditch et al. [30]). In interacting binaries, irradiation can alter stellar structure [31,32] and drive cyclical mass-transfer (MT) transitions [29,33–35]. Crucially, irradiation-driven wind enhancement [36–38] provides a direct mechanism to amplify MB [39,40]. In extreme cases, persistent irradiation can evaporate the companion entirely, as observed in "black widow" pulsars [41–45].

While saturated magnetic braking (SMB) has been applied in isolation to LMXB evolution, its failure to resolve key observational puzzles has hindered its widespread adoption [11]. We argue that this limitation likely stems from the omission of irradiation-driven winds, a physical process that can significantly enhance AM loss. To address this shortcoming, we couple these two mechanisms in our Saturated Magnetic Braking and Irradiation-Driven Wind (SMB + IDW) model. In this framework, irradiation-driven winds directly modulate the efficiency of MB, leading to a self-consistent AM loss prescription.

## 2. Methods

### 2.1. The stellar evolution code

We computed binary evolution using the Modules for Experiments in Stellar Astrophysics (MESA) code [46,47]. Each system was initialized with a neutron star (NS) and a main-sequence (MS) companion of solar metallicity [48] ($Z$ = 0.0142). The microphysics in our simulations was governed by standard prescriptions: opacities were derived from the OPAL tables [49], complemented by the low-temperature tables of Ferguson et al. [50]; the equation of state (EOS) was constructed via a blend of the OPAL [51], HELM [52], PC [53], and SCVH [54] prescriptions; and nuclear reaction rates were primarily sourced from the JINA REACLIB compilation [55], with supplemental rates from Angulo et al. [56] (NACRE) for reactions not included in the primary source.

Convection was treated using the mixing-length theory of Böhm-Vitense [57] as implemented by Cox et al. [58]. The mixing-length parameter, $\alpha_{\rm MLT}$, was calibrated against solar properties, yielding a value of 2 within the typical range of 1.5–2.0 [59]. No convective overshooting was included in our models. For stellar winds, we employed empirically calibrated prescriptions: the Reimers [60] formula was applied during the MS and red-giant phases, while the Bloecker [61] relation was utilized for the asymptotic giant branch phase.

Our parameter space exploration employed the following configurations. The initial NS mass was set to $1.3\,M_\odot$, with the donor mass ranging from 0.8 to $4.0\,M_\odot$ in increments of $0.2\,M_\odot$. The initial orbital period was sampled uniformly in logarithmic space over $-0.5 \leqslant \log(P_{\rm orb,i}/{\rm day}) \leqslant 3.0$, with a step size of $\Delta \log P_{\rm orb,i} = 0.05$.

### 2.2. Mass transfer and irradiation wind models

We calculated the effective Roche-lobe (RL) radius of the secondary star using the formula from Eggleton [62]. Mass transfer was modeled using a modified Kolb et al. [63] prescription, which accounted for extended atmospheres [64] and permitted optically thick Roche-lobe overflow (RLO). We adopted an Eddington-limited accretion prescription for the NS: $\dot{M}_{\rm NS} = \min(\dot{M}_{\rm Edd}, \dot{M}_{\rm tr})$, where $\dot{M}_{\rm tr}$ is the binary MT rate. Super-Eddington MT was assumed to be expelled as an isotropic wind from the vicinity of the NS, carrying the specific orbital AM:

$$\dot{J}_{\rm ML,Super-Edd} = -\left(\frac{M_2 a}{M_{\rm total}}\right)^2 (\dot{M}_{\rm tr} - \dot{M}_{\rm Edd})\Omega_{\rm orb}, \tag{1}$$

where $M_2$ and $M_{\rm total}$ are the donor mass and total mass of the binary, $a$ is the binary separation, and $\Omega_{\rm orb}$ is the orbital angular velocity. This prescription was justified by the fact that alternative mass-loss channels did not qualitatively alter our conclusions. For instance, mass loss via the $L_1$ point led to specific AM loss comparable to that removed in the isotropic-wind case, resulting in a negligible difference in the net orbital evolution [65]. In contrast, significant outflow through the $L_2$ point could have formed a circumbinary disk and driven unstable MT, potentially leading to common-envelope (CE) evolution [66]. However, such a configuration generally required the companion star to fill its outer Lagrangian lobe, a condition not commonly observed in the LMXBs studied here, nor was it necessary within our evolutionary framework. Therefore, the isotropic wind model represented a well-motivated and self-consistent simplification for our computational approach.

During accretion, the infalling material converted gravitational potential energy into kinetic energy. Upon impacting the NS surface, this energy thermalized and was radiated as intense X-rays, with luminosity $L_{\rm X} = GM_{\rm NS}\dot{M}_{\rm NS}/R_{\rm NS}$, where $G$ is the gravitational constant and $R_{\rm NS}$ is the radius of the NS. This radiation illuminated and heated the outer layers of the donor star. Theoretically, this irradiation was capable of driving a substantial stellar wind when the net heating rate $A$ — defined as external heating minus radiative cooling — became positive. This condition was typically met in the low-density outer atmospheric layers, where heating processes such as Compton scattering and photoionization dominated over radiative cooling [36,37]. Once the local pressure fell below a critical value $P_0$ (or equivalently, the density dropped below a critical threshold), heating overcame the local gravitational binding, allowing gas to become unbound and drive a wind [36,37].

We linked the irradiation-driven mass loss to the accretion power via a simple energy argument. The material in the wind was assumed to become only marginally unbound [37,39], so that its specific kinetic energy at infinity was negligible. The minimum power per unit mass required to lift gas from the donor surface was therefore the gravitational binding energy $GM_2/R_2$ (where $R_2$ is the donor star radius). The power actually available to drive the wind was expressed as a fraction of the X-ray luminosity that intercepted the donor and overcame atmospheric cooling. Denoting the nominal (maximum) accretion luminosity by $L_{\rm X} = GM_{\rm NS}\dot{M}_{\rm NS}/R_{\rm NS}$, the wind-driving power was written as $f_x f_\Omega f_\epsilon L_{\rm X}(R_2^2/4a^2)$, where the three dimensionless factors are:

$f_x$: the radiative efficiency of accretion, i.e., the fraction of the liberated gravitational energy that was radiated as X-rays. By definition $f_x \leqslant 1$, with $f_x = 1$ corresponding to the nominal $L_{\rm X}$ used here.

$f_\Omega$: a geometric correction to the isotropic dilution factor $R_2^2/(4a^2)$ (assuming $a \gg R_2$), accounting for anisotropic emission and possible occultation by the accretion disk.

$f_\epsilon$: the net fraction of the absorbed radiative energy that drove the wind after subtracting radiative cooling and other atmospheric losses. Detailed calculations by Tavani et al. [37] placed this wind-driving efficiency in the range 0.001–0.1.

Equating the required binding power with the supplied mechanical power,

$$\dot{M}_{\rm wind,irr}\frac{GM_2}{R_2} = -f_x f_\Omega f_\epsilon L_{\rm X}\frac{R_2^2}{4a^2}, \tag{2}$$

yielded the mass-loss rate

$$\dot{M}_{\rm wind,irr} = -f\frac{M_{\rm NS}R_2^3\dot{M}_{\rm NS}}{4a^2R_{\rm NS}M_2}, \tag{3}$$

where the total efficiency was defined as

$$f = f_x f_\Omega f_\epsilon. \tag{4}$$

Since all three sub-factors were individually smaller than unity, their product $f$ was naturally a very small number. Previous work estimated the overall effective efficiency in the range $10^{-5}$–$10^{-1}$ [37,39,40]. In this study, we adopted a conservative interval $f = 10^{-5}$–$10^{-3}$, reflecting the compounded uncertainties in $f_x$,$f_\Omega$, and $f_\epsilon$.

A critical aspect of our binary evolution calculation was the criterion for unstable MT leading to a CE phase. In an RLO NS binary, the transferred material formed an accretion disk around the NS. Sufficiently high MT rates led to matter accumulation around the NS, potentially forming an optically thick, bloated cloud that engulfed much of the disk. Radiation became trapped within the so-called photon-trapping radius, defined as $R_{\rm trap} = (R_{\rm S}/2)(\dot{M}_{\rm tr}/\dot{M}_{\rm Edd})$ [67]. Here, $R_{\rm S}$ denoted the Schwarzschild radius for a black hole (BH); for a NS, we set $R_{\rm S} \simeq R_{\rm NS}/2$. We assumed the system entered a CE-like phase if $R_{\rm trap}$ exceeded the NS's RL radius $R_{\rm L,NS}$, or equivalently, if the MT rate surpassed the critical value $\dot{M}_{\rm trap} = 2\dot{M}_{\rm Edd}(R_{\rm L,NS}/R_{\rm S})$ [68,69]. Under other conditions, RLO via the $L_1$ point might have been insufficient to strip the donor's expanding envelope, allowing the donor star to expand beyond its $L_2$ lobe. Mass loss through the $L_2$ point could extract substantial AM from the binary, leading to rapid orbital contraction and potentially triggering CE evolution (e.g., Refs. [66,70]). However, studies of MT stability in binaries with convective giant donors indicated that systems might avoid CE evolution even after $L_2$ overflow begins [71]. In our simulations, we assumed that CE evolution was triggered if either of the following two criteria was met: (1) the MT rate exceeded the trapping limit: $\dot{M}_{\rm tr} > \dot{M}_{\rm trap}$ [72]; or (2) the system experienced $L_2$ overflow and the MT rate was dynamically high, specifically when both $R_2 > R_{L_2}$ [66] and $\dot{M}_{\rm tr} > 0.02\,M_2/P_{\rm orb}$ were satisfied [71]. The computation was terminated upon CE entry or after 10,000 models.

### 2.3. Magnetic braking model

We utilized a standard MB prescription for LMXBs, based on the Skumanich-type $\Omega^3$ scaling proposed by Skumanich [4] and adapted to binary evolution by Verbunt et al. [5] and Rappaport, Verbunt, Joss [6] (RVJ83). This prescription was expressed as:

$$\dot{J}_{\rm RVJ83} = -3.8\times10^{-30}M_2R_\odot^4\left(\frac{R_2}{R_\odot}\right)^{\gamma_{\rm mb}}\Omega^3\,{\rm erg}, \tag{5}$$

where $\Omega$ is the donor's angular rotation velocity, and $\gamma_{\rm mb}$ is a dimensionless parameter typically ranging from 0 to 4. We assumed that the donor star's rotation was tidally synchronized with the orbit. This assumption was well justified. First, observations by Lurie et al. [73] showed that MS binaries with orbital periods shorter than 10 d were all fully synchronized, with only a weak dependence on mass. Second, according to the tidal study of compact binaries by Repetto et al. [74], the synchronization timescale for typical compact-object (NS/BH) LMXBs was shorter than the MS lifetime of the low-mass companion by at least two orders of magnitude (see their Fig. 4); in other words, tidal synchronization was rapid in a relative sense. Furthermore, for binaries with longer orbital periods, even if synchronous rotation was assumed, MB had a negligible impact on the orbit before RL filling. This was demonstrated in Fig. 9 of Deng et al. [11], where the same assumption of synchronous rotation was adopted and the orbit remained essentially unchanged prior to RLO. Finally, once the donor overfilled its RL, the transferred material formed an accretion disk around the compact object. This disk exchanged AM with the binary via Lindblad resonances [75], which clearly drove a rapid synchronization of the stellar spin with the orbital motion as well.

However, in rapidly rotating stars, the magnetic field was expected to saturate, breaking the simple $\Omega^3$ scaling. To account for this, we adopted the saturated MB prescription from Matt et al. [24] (Matt15). This model was grounded in the physical picture that magnetized winds carry away AM [1,3], and it incorporated a mass-dependent saturation threshold that better reproduced the spin-down of Sun-like stars in stellar population studies.

The model's key innovation was its explicit dependence on the Rossby number $Ro = P_{\rm rot}/\tau_c$ (the ratio of rotation period to convective turnover time), which robustly governs stellar magnetic activity and its saturation [76–78]. It featured a two-part torque law that distinguished saturated from unsaturated magnetic regimes:

1. Saturated Regime ($Ro \leqslant Ro_\odot/\chi$): for rapid rotators, where magnetic activity saturates [76,79], the torque scales linearly with angular velocity:

$$T_{\rm sat} = -T_0\chi^p\left(\frac{\Omega}{\Omega_\odot}\right). \tag{6}$$

2. Unsaturated Regime ($Ro > Ro_\odot/\chi$): for slower rotators, where magnetic activity scales with $Ro$ [77,78], the torque follows a steeper dependence:

$$T_{\rm unsat} = -T_0\left(\frac{\tau_c}{\tau_{c,\odot}}\right)^p\left(\frac{\Omega}{\Omega_\odot}\right)^{p+1}. \tag{7}$$

Here, $\chi$ defines the critical Rossby number for saturation, $\tau_{c,\odot} = 12.9\,{\rm d}$ and $\Omega_\odot = 3\times10^{-6}\,{\rm rad\,s^{-1}}$ are the solar convective turnover time and angular velocity, respectively. We adopted the standard values $\chi = 10$ and $p = 2$; the latter recovered the canonical Skumanich-type spin-down ($T \propto \Omega^3$) for solar-mass stars [4].

The model's key strength was its empirically calibrated normalization $T_0 \propto R_2^{3.1}M_2^{0.5}$ [24], a scaling critically responsible for its predictive success. This specific mass dependence accounted for two seemingly contradictory trends: the prolonged fast rotation of low-mass stars in their youth and their eventual emergence as the slowest rotators on the old cluster sequence [80,81]. The model further demonstrated its predictive capability by reproducing subtle features in the Kepler rotation period distribution, including the observed upper envelope and the sharp lower boundary that coincides with the theoretical saturation sequence [24,82]. By self-consistently linking stellar structure, magnetic field generation, and wind properties, it provided a coherent framework for interpreting stellar rotation across mass and age.

The efficiency of MB was fundamentally coupled to stellar wind mass loss. In LMXBs, X-ray irradiation from the accretor substantially enhanced the mass-loss rate ($\dot{M}_{\rm wind}$) of the companion star compared to isolated stars [36–38]. Building on the model of Justham et al. [39], in which the MB-driven AM loss rate scales as $\dot{J}_{\rm mb} \propto \left(\dot{M}_{\rm wind}\right)^{0.5}$, we incorporated irradiation by defining an enhancement factor:

$$\eta = \left(\frac{\dot{M}_{\rm wind,irr} + \dot{M}_{\rm wind}}{\dot{M}_{\rm wind}}\right)^{0.5}, \tag{8}$$

where $\dot{M}_{\rm wind,irr}$ and $\dot{M}_{\rm wind}$ are the irradiation-driven and intrinsic wind mass-loss rates, respectively. The total AM loss rate under irradiation was then $\dot{J}_{\rm mb,irr} = \eta \cdot \dot{J}_{\rm mb}$.

Finally, integrating this with the saturated MB formalism [24] yielded our complete expression:

$$\dot{J}_{\rm SMB,IDW} = \begin{cases} -\eta \cdot \tau_0 P_{{\rm rot},-3}\, \tau_{c,2}\, R_{3.1}\, M_{0.5}, & P_{\rm rot} \geqslant P_{\rm crit} \\ -\eta \cdot \tau_0 P_{{\rm rot},-1}\, P_{{\rm crit},-2}\, \tau_{c,2}\, R_{3.1}\, M_{0.5}, & P_{\rm rot} < P_{\rm crit} \end{cases}, \tag{9}$$

where we defined the dimensionless parameters: $P_{{\rm rot},-3} = (P_{\rm rot}/1\,{\rm day})^{-3}, P_{{\rm rot},-1} = (P_{\rm rot}/1\,{\rm day})^{-1}, P_{{\rm crit},-2} = (P_{\rm crit}/1\,{\rm day})^{-2}$, $\tau_{c,2} = (\tau_c/\tau_{c,\odot})^2, R_{3.1} = (R_2/R_\odot)^{3.1}$, and $M_{0.5} = (M_2/M_\odot)^{0.5}$. Here, the normalization constant was $\tau_0 = 1.38 \times 10^{35}$ erg. The critical period for magnetic saturation was $P_{\rm crit} = 0.1\,P_\odot(\tau_c/\tau_{c,\odot})$, where $P_\odot = 24.23$ d is the Solar spin period. Importantly, the irradiation-driven enhancement in this model did not increase indefinitely with the MT rate. Due to the Eddington limit on accretion luminosity, both the irradiating flux and its coupling to MB saturated at high transfer rates, providing a natural physical upper bound and preventing unphysical runaway feedback.

### *2.4. The binary population synthesis code*

We performed population synthesis simulations of NS-MS binary formation using the Binary Star Evolution (BSE) code [83]. We adopted a constant star formation rate of $3\,M_\odot\,{\rm yr}^{-1}$ over 12 Gyr. For primordial binaries, primary masses followed a Kroupa et al. [84] initial mass function between $5$–$40\,M_\odot$, secondaries were assigned via a uniform mass-ratio distribution [85], and orbital separations were sampled logarithmically from 3 to $10^4\,R_\odot$ [86], all starting in circular orbits.

The stellar wind models implemented in BSE and MESA were tailored to their respective mass domains. Within the BSE framework, which tracks massive-star populations, we applied the wind prescription of Nieuwenhuijzen et al. [87], with OB stars and stripped helium stars following the theoretical relations from Vink et al. [88,89]. During RLO, we assumed half of the transferred mass was accreted by the secondary and applied the criterion of Shao et al. [90] to determine MT stability. Unstable MT led to a CE phase, which we modeled using the standard energy formalism [91] with an ejection efficiency $\alpha_{\rm CE} = 1$.

NS formation followed the rapid explosion model of Fryer et al. [92], which also provided the criteria to distinguish between core-collapse supernovae (CCSNe) and electron-capture supernovae (ECSNe). Natal kicks were drawn from Maxwellian distributions with $\sigma_{\rm CCSN} = 265\,{\rm km\,s^{-1}}$ for CCSNe and $\sigma_{\rm ECSN} = 20\,{\rm km\,s^{-1}}$ for ECSNe [93–95].

The binary population synthesis described above served a specific purpose within our overall methodology. Starting from $10^8$ primordial binaries, we evolved each system through all relevant physical processes until a NS-MS binary was formed. The primary outputs of this stage were the masses, orbital periods, and eccentricities of the resulting NS-MS binaries, together with the birth rate (formation weight) associated with each system. These parameters and weights defined the initial conditions for the subsequent long-term evolution of LMXBs.

The detailed evolutionary tracks of individual NS-MS binaries were then computed using the MESA code. For a given initial donor mass, initial orbital period, and NS mass drawn from the population synthesis, MESA provided the time-resolved evolution of the MT rate, donor mass and radius, orbital period, and other quantities of interest. By combining the population synthesis results (which provided the statistical distribution of initial conditions and formation rates) with the MESA tracks (which provided the physical evolution along each track), we could compute observable statistical properties of the LMXB population and of their descendants — the recycled pulsar binaries. Specifically, the number of systems in a given observational bin was obtained by summing over all evolutionary tracks, weighting each by its formation rate and the time spent in that bin.

## 3. Results

### *3.1. A representative evolutionary pathway and model comparison*

Fig. 1 shows the evolutionary trajectory of a representative system under the SMB + IDW model: a $1.3\,M_\odot$ NS with a $1.2\,M_\odot$ MS companion (initial orbital period $P_{\rm orb} = 2.24$ d, irradiation efficiency $f = 10^{-4}$). The system starts detached (Phase 1). The short orbital period justifies the assumption of tidal synchronization between the orbit and the companion's rotation. This synchronization allows MB to directly extract orbital AM.

Orbital decay induced by MB leads to RL filling by the donor at 4.23 Gyr, initiating MT (Phase 2). The resulting accretion luminosity, $L_{\rm X}$, irradiates the donor. This heats the donor's outer atmosphere until net heating overcomes local gravitational binding, driving an enhanced thermal wind. Coupled to the donor's magnetic field, this wind directly amplifies AM loss via MB, establishing a self-regulating feedback loop: increased MT raises $L_{\rm X}$, which strengthens the wind, further amplifying the MB torque and modifying the orbit and subsequent transfer rate. The loop saturates as $L_{\rm X}$ approaches the Eddington limit.

After 2.58 Gyr of MT, the donor's hydrogen envelope is stripped, leaving a $\sim 0.139\,M_\odot$ helium core. The system becomes a 5.5 h MSP-helium WD binary (Phase 3). Gravitational wave emission subsequently shrinks the orbit over 4.02 Gyr until, at 10.83 Gyr, the companion refills its RL, forming an UCXB with $P_{\rm orb} < 1\,{\rm h}$ (Phase 4) — a potential source of gravitational waves [96].

We compare three MB prescriptions in Fig. S1 (online): the traditional RVJ83 law, the Matt15 model, and our SMB + IDW model. Prior to RLO, SMB + IDW is identical to Matt15 because no accretion-powered X-rays are produced to irradiate the donor; both produce much weaker braking than RVJ83. Consequently, in the SMB + IDW and Matt15 models, the donor star evolves off the MS and forms a helium core before filling its RL, whereas the strong RVJ83 braking drives such rapid orbital contraction that the donor overfills its RL while still on the MS, never developing a degenerate helium core.

Once MT begins, the SMB + IDW model diverges from Matt15: accretion-powered X-ray irradiation drives an enhanced stellar wind from the donor, which amplifies MB. The AM loss efficiency is illustrated for three assumed irradiation efficiencies, $f = 10^{-5}, 10^{-4}$ (our reference value), and $10^{-3}$. At $f = 10^{-5}$, the braking strength increases by about an order of magnitude relative to the saturation-only case (Matt15), reaching levels comparable to the traditional RVJ83 prescription. For $f = 10^{-4}$, the braking strength rises by more than half an order of magnitude but less than a full order; for $f = 10^{-3}$, the additional gain is only a factor of two to three. We note that the degree of irradiation-boosted MB enhancement decays as the donor mass decreases, closely following the decline of the MT rate along the evolutionary track (Fig. S1b online). Nevertheless, even when the donor has lost a large fraction of its envelope and is reduced to $\sim 0.5M_\odot$, the braking strength in the reference model ($f = 10^{-4}$) remains more than an order of magnitude higher than that of Matt15. The overall saturation effect intensifies with the highest irradiation efficiencies, which push the MT rate to near or past the Eddington limit, thereby saturating the X-ray luminosity and quenching additional wind enhancement. For our reference model ($f = 10^{-4}$), the MT rate is enhanced by up to a factor of $\gtrsim 20$ compared to Matt15 and is somewhat less than an order of magnitude higher than the RVJ83 case.

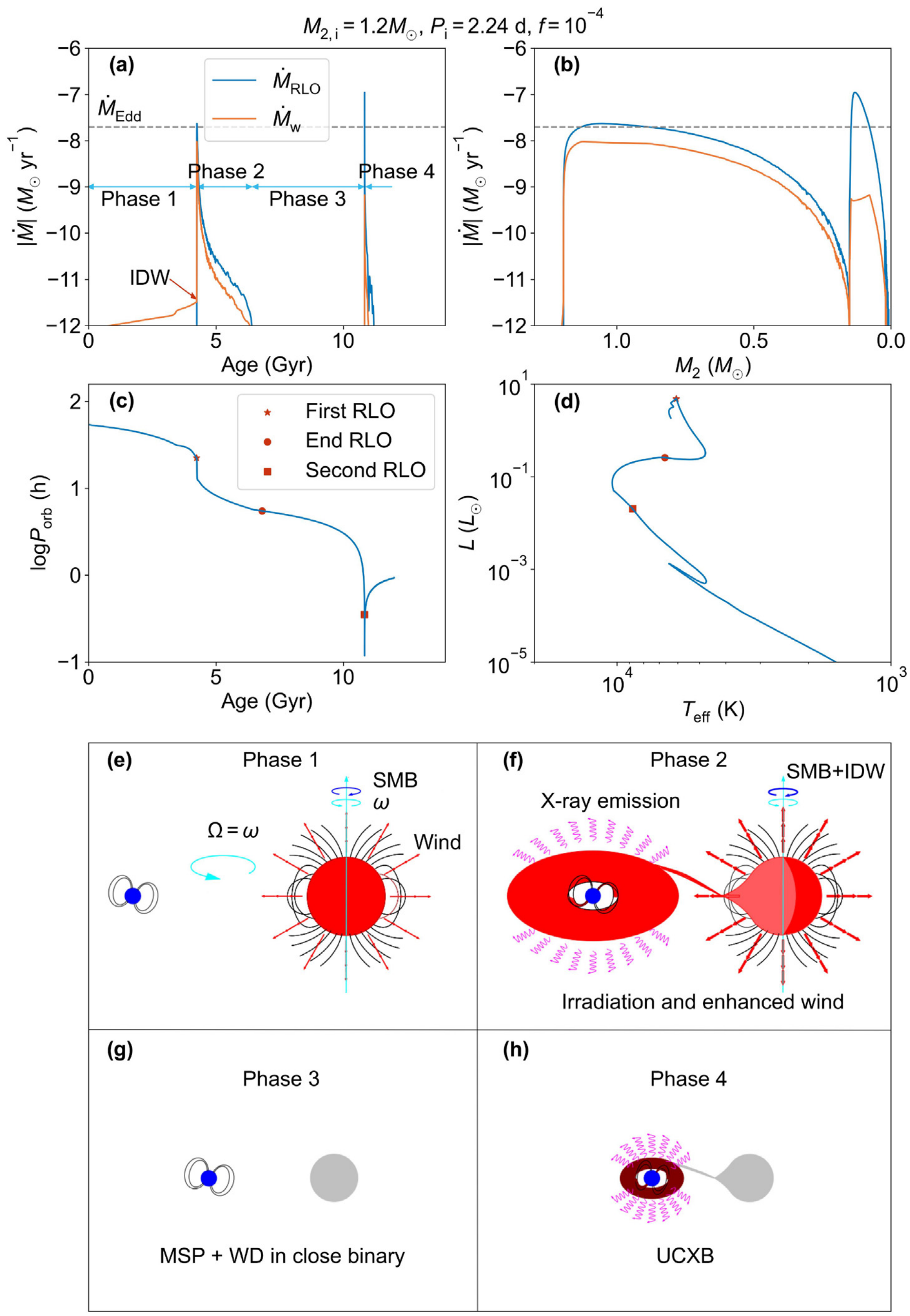

**Fig. 1.** Schematic representation of the evolution of a LMXB within the SMB + IDW model. The system is initialized with a $1.3M_\odot$ NS and a $1.2M_\odot$ MS companion, with an initial orbital period of 2.24 d and an irradiation efficiency $f = 10^{-4}$. The panels (a) and (b) illustrate MT and wind loss processes: the blue solid line denotes the RLO MT rate, the orange solid line represents the companion's wind mass-loss rate, and the grey dashed line indicates the Eddington accretion rate of the NS. The panels (c) and (d) display the orbital period evolution and the donor's Hertzsprung-Russell diagram, with key epochs marked by a red star (onset of RLO), a red circle (RL detachment), and a red square (second RLO). The panels (e–h) depict schematic diagrams of the four characteristic evolutionary stages of the system.

These differences in braking directly shape the orbital evolution. In SMB + IDW, the persistently strong irradiation-driven braking sustains AM loss after RLO, driving continued orbital contraction that ultimately leads to a UCXB composed of a NS and a helium WD. In contrast, the Matt15 model — lacking irradiation feedback — experiences weaker post-RLO braking; MT dominates the orbital response and the binary expands, ending as an ordinary pulsar-helium WD binary with an orbital period of less than 10 d. The SMB + IDW model thus uniquely satisfies both conditions for UCXB formation: it allows a degenerate helium core companion to develop during the pre-RLO phase, and it provides the post-RLO AM loss mechanism required to shrink the orbit into the ultracompact regime.

### *3.2. A unified framework for LMXB Diversity: accretion rates, the "fine-tuning" problem, and anomalous orbital expansion*

Having demonstrated the evolutionary mechanism for a single system, we evaluate the model across a broad parameter space. Fig. 2 shows evolutionary trajectories for NS LMXBs under different irradiation efficiencies. Without irradiation (Fig. 2a), the models reproduce the well-known bifurcation period near 10 h [7]. Systems with longer periods expand continuously, whereas those with shorter periods contract, producing a pronounced gap in the theoretical period distribution. The calculated MT rates in this case are 1–2 orders of magnitude below the observed values for persistent LMXBs, and the formation of ultracompact binaries (periods <1 h) is strongly suppressed — results that are consistent with earlier studies [7,11,97].

With irradiation included (Fig. 2b), the predictions change substantially. Higher irradiation efficiencies yield significantly higher MT rates, enabling the models to reproduce the full observed range of accretion rates. Moreover, the formation of NS UCXBs is successfully reproduced.

A quantitative comparison with the observed NS LMXB population is provided in Table S1 (online). The SMB + IDW model reproduces all key observed parameters across the range of irradiation efficiencies considered. The systematic difference between transient and persistent sources is captured by the efficiency parameter $f$: lower values ($f \sim 10^{-5}$–$10^{-4}$) correspond to transients, and higher values ($f \sim 10^{-4}$–$10^{-3}$) to persistent systems. This dependence suggests that efficient irradiation-driven winds require sustained X-ray illumination, a condition less consistently met during the episodic outbursts of transient sources.

A persistent challenge in LMXB evolution is the "fine-tuning" problem: conventional MB models require initial parameters to lie within an extremely narrow range to produce the observed populations of tight-orbit MSP-helium WD binaries [8]. Our results demonstrate that the SMB + IDW model naturally alleviates this problem.

Fig. 3 tracks the orbital period evolution of LMXBs over time, with the grey band marking the 2–9 h period window. The colour scale indicates the donor's hydrogen abundance ($X$): $X > 0.6$ for the MS phase and $X < 0.05$ for a WD with a thin H envelope. A key visual result is that evolutionary trajectories enter and cross this shaded band only when irradiation is included; specifically, the red lines (H  < 0.05) correspond to helium WD companions.

This visual contrast reflects a clear quantitative outcome. Without X-ray irradiation, none of the systems in our parameter grid evolve into MSP-helium WD binaries within the 2–9 h range. When irradiation is active, such systems form successfully. These resulting binaries will continue to contract via gravitational-wave radiation within a Hubble time, eventually causing the helium WD to refill its RL and form an ultracompact NS-helium WD X-ray binary [8].

The parameter space leading to these ultracompact X-ray binaries is shown in Fig. S2 (online) for different irradiation efficiencies. In the absence of irradiation, no systems within the explored grid form such binaries. As the irradiation strength increases, the range of viable initial parameters widens progressively, indicating that irradiation reduces the fine-tuning required in conventional models. Furthermore, the bifurcation period — defined as the orbital period at the onset of RLO that separates evolutionary channels — increases with stronger irradiation.

While traditional models struggle to account for the anomalous orbital expansion observed in some LMXBs [14], the SMB + IDW model successfully explains this phenomenon. The NS LMXB 2A1822–371 ($P_{\rm orb} = 5.57$ h) [98] exemplifies this behavior: as a short-period binary, standard MB predicts orbital contraction, whereas measurements indicate rapid expansion [40]. Explaining this expansion with conventional models would require a super-Eddington MT rate. Xing et al. [40] applied the irradiation + strong-field MB prescription proposed by Justham et al. [39] to the peculiar source 2A1822–371, an assumption requiring a donor magnetic field strength that is rare among low-mass stars. Although formulated differently, the SMB + IDW model yields consistent results for this system. As shown in Fig. 4, an SMB + IDW evolutionary track matches the observed expansion rate, orbital period, and donor mass of 2A1822–371 [99–101].

### *3.3. Predictions for the orbital period distribution of binary millisecond pulsars*

Having demonstrated that the SMB + IDW model resolves key puzzles in LMXB evolution, we now examine its predictions for the orbital period distribution of the resulting binary MSPs. To perform a quantitative comparison with observations, we construct theoretical orbital period distributions using initial conditions derived from binary population synthesis [83]. We select observed systems with orbital eccentricity $e < 0.01$, consistent with efficient orbital circularization during the LMXB phase [102]. To ensure a physically meaningful comparison, we divide the observational sample into two subsets: (i) all pulsar binaries with $P_{spin} < 30$ ms, and (ii) the cleaner subset of MSP-helium WD systems ($P_{spin} < 10$ ms), which are predominantly descended from LMXBs.

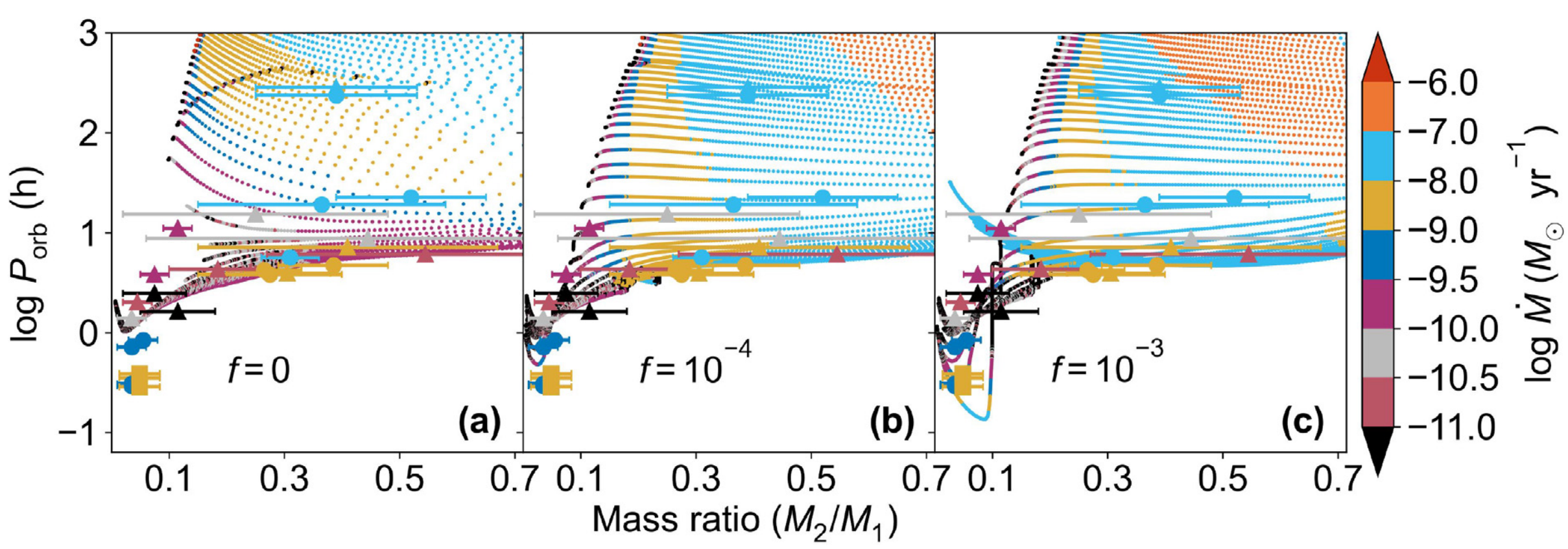

**Fig. 2.** Evolutionary trajectories of NS LMXBs under the SMB + IDW model. Each panel corresponds to a donor initial mass of $1.0 M_\odot$ and displays results for different irradiation efficiencies: $f = 0$ (panel a), $f = 10^{-4}$ (panel b), and $f = 10^{-3}$ (panel c). The colour of each track encodes the MT rate, as indicated by the colour bar on the far right. Symbols with error bars denote observed systems: circles, squares, and triangles correspond to persistent LMXBs, globular cluster (GC) LMXBs, and transient LMXBs, respectively.

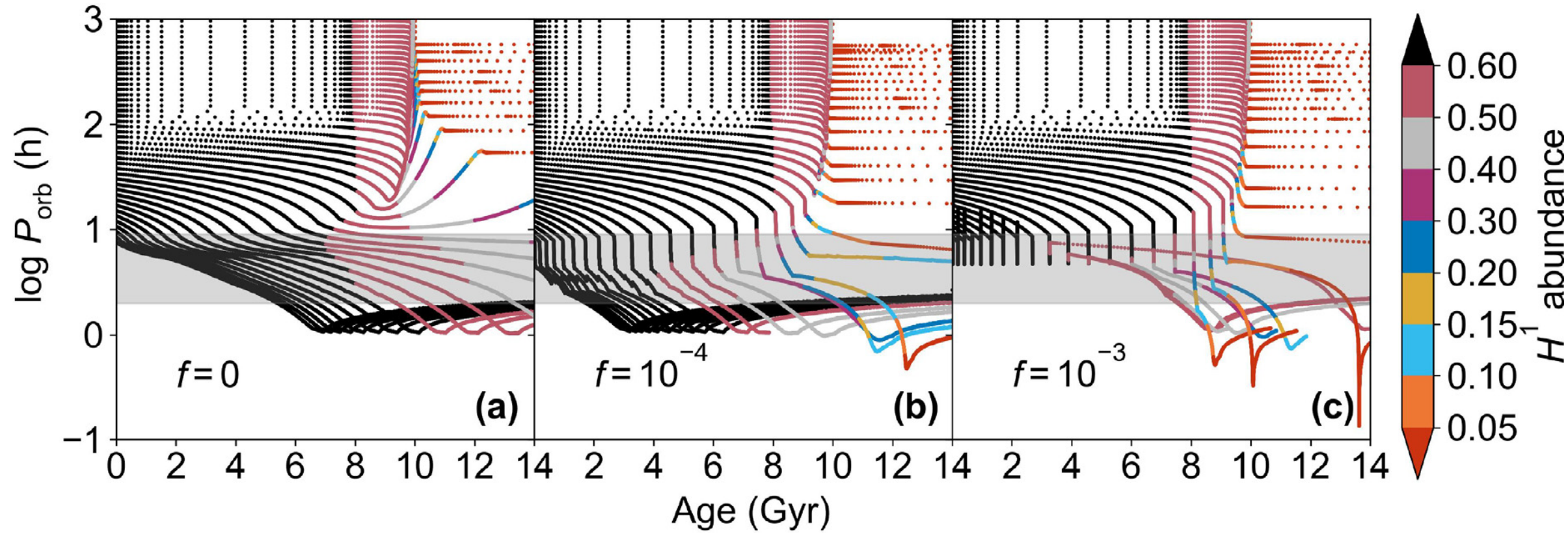

**Fig. 3.** Orbital period evolution over time for NS LMXBs with a $1.0M_\odot$ donor, under different irradiation efficiencies: $f = 0$ (panel a), $f = 10^{-4}$ (panel b), $f = 10^{-3}$ (panel c). Colour indicates the hydrogen abundance of the donor star.

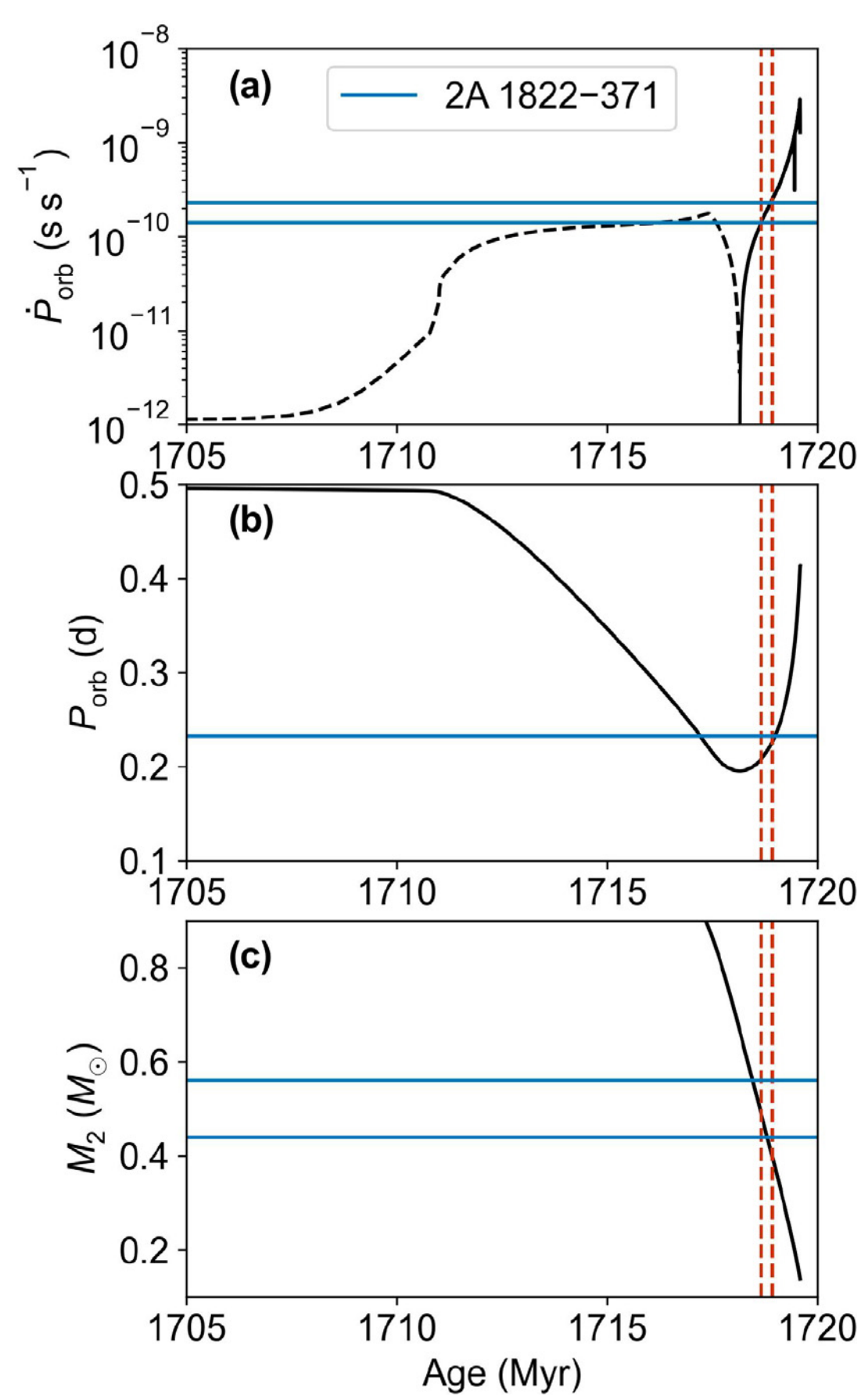

**Fig. 4.** Binary evolution model for 2A 1822−371. The system is initialized with a $1.3M_\odot$ NS and a $1.2M_\odot$ donor star at an initial orbital period of 1 d and an irradiation efficiency $f = 0.005$. Panels (a–c) show the orbital period derivative, orbital period, and donor mass as functions of time. In panel (a), the black dashed and solid lines denote orbital contraction and expansion, respectively. Blue lines denote observational data. The red dashed band (1718.66–1718.93 Myr) highlights the epoch when the modelled orbital expansion rate matches the observed rate.

Fig. 5 compares the theoretical and observed orbital period distributions for both subsets. The SMB + IDW model reproduces the overall observed range ($0.1 \lesssim P_{orb} \lesssim 1000$ d) and the prominent peak near 10 d in both samples. Focusing on the cleaner MSP-helium WD sample, gaps exist in the observed distribution at 2.4–4.3 d and ~25–50 d [103]. The latter corresponds to the historically identified "Camilo" gap (~13–68 d; [104,105]). While subsequent discoveries have populated this broad interval, a deficit persists and is more precisely localized between 16.3 and 28.8 d (Fig. S3 online).

The theoretical distribution depends on the irradiation efficiency $f$. The short-period gap, associated with the bifurcation period [7], is reproduced for $f = 0 - 10^{-4}$. The location and shape of the tens-of-days gap vary with $f$; a value of $f \sim 10^{-3}$ produces a theoretical deficit that aligns most closely with the observed gap at 16.3–28.8 d. A physical origin for this gap is identified within the SMB + IDW framework (Fig. S3 online). Systems with donors of $M_{2,\mathrm{i}} \sim 1.6$–$1.8 M_\odot$ and initial periods of several to tens of days undergo unstable MT under the criteria described in Section 2.2, creating a broad intrinsic period gap. One possible driver of this instability, not explored in detail here, is the rapid expansion accompanying the first dredge-up during the red giant phase, which may destabilize MT even in otherwise stable binaries. Irradiation subsequently shifts this gap systematically toward shorter periods. This mechanism offers a new explanation that differs from those invoking a bifurcation period [106] or bump-induced detachment [107], establishing a self-consistent link between binary instability and the observed population structure.

Fig. 5a further reveals that the theoretical orbital distribution of pulsar binaries fails to adequately reproduce the observed peak at orbital periods of ~2–5 h. This discrepancy likely stems from multiple factors. We note that our treatment simplifies the orbital and mass parameters of pulsar binaries that emerge from CE evolution using the Webbink [91] energy formalism, subsequently evolving these systems via gravitational radiation [108]. However, the conditions for entering CE evolution [109–111] (MT instability), the CE process [112], the state of the helium core after successful envelope ejection [113], and the potential formation of a circumbinary disk from fallback material [114–116] can all profoundly affect the resulting orbital distribution. These uncertainties are likely the primary cause of the tension between theory and observations.

Additionally, the results in Fig. 5 show a significant fraction of binaries with orbital periods between ~200 and 1000 d, which is in stark contrast to the observed distribution. This likely arises because wide-orbit binaries are more susceptible to unstable MT, leading to CE evolution and subsequent merger or the formation of a close binary, rather than surviving as wide systems. Another contributing factor is selection bias. In Fig. S4 (online), we present the distributions of orbital period, spin period, and surface magnetic field (B) for 334 pulsar binaries compiled from the ATNF Pulsar Catalogue [117]. This figure indicates that among binaries with orbital periods below ~200 d, the vast majority are fully recycled MSPs with spin periods $< 10$ ms and surface magnetic fields around $10^8$ G. In contrast, for systems with orbital periods $> 200$ d, no fully recycled MSPs with a spin period $< 10$ ms and a magnetic field $< 10^9$ G are found; instead, they are predominantly mildly recycled or non-recycled pulsars with magnetic fields ranging from a few times $10^9$ G to a few times $10^{12}$ G and spin periods from tens of milliseconds to nearly one second. This pattern is consistent with theoretical expectations: in wide binaries, the donor star typically fills its RL at a late evolutionary stage, resulting in an MT timescale significantly shorter than that in close binaries. Evidently, due to magnetic dipole radiation, fully recycled, low-B MSPs enjoy significantly longer observational lifetimes than mildly recycled and non-recycled, high-B pulsars [118]. Consequently, a substantial population of wide-orbit binary pulsars may have already spun down and become radio-quiet, evading detection in radio surveys. Recently, the Gaia mission has discovered over twenty NS-MS binaries with orbital periods of 200–1000 d through optical surveys [119], which may serve as observational evidence supporting this hypothesis.

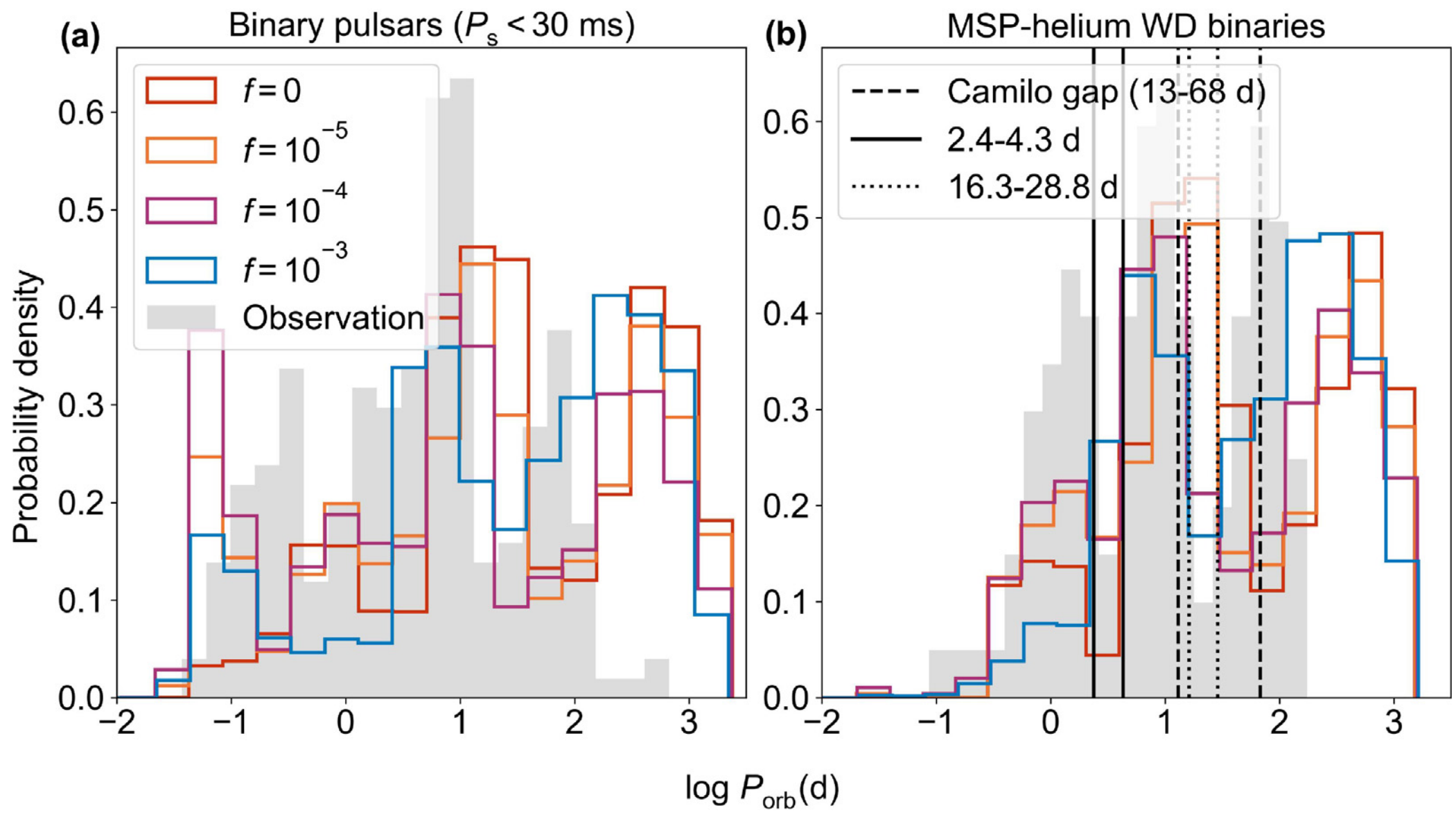

**Fig. 5.** Comparison of theoretical and observed orbital period distributions for pulsar binaries. Theoretical distributions (coloured curves, colour-coded by irradiation efficiency: red $f = 0$, orange $f = 10^{-5}$, purple $f = 10^{-4}$, blue $f = 10^{-3}$) are compared to observed histograms (grey shading). Panel (a): all simulated pulsars with accreted mass $> 0$ and observed pulsars with spin period $< 30$ ms and eccentricity $< 0.01$. Panel (b): simulated MSP-helium WD binaries (accreted mass $> 0.05 M_\odot$) and observed MSP-helium WD systems (spin period $< 10$ ms, eccentricity $< 0.01$). Vertical black lines mark key period gaps in the observed population: the dashed line denotes the historically identified "Camilo gap" (~13–68 d, [104]), the dotted line highlights a persistent, pronounced gap at 16.3–28.8 d, and the solid line indicates another significant gap at 2.4–4.3 d.

## 4. Discussion

### *4.1. Extension to black-hole LMXBs and cataclysmic variables*

Previous studies have proposed distinct MB prescriptions for BH and NS LMXBs [18], despite the intrinsic similarity between these two classes of systems. This tension calls for a unified framework and serves as a key motivation for the present work. Here we demonstrate that the SMB + IDW framework, originally developed for NS LMXBs, performs similarly well for BH systems. Fig. S5 (online) shows evolutionary pathways for an $8 M_\odot$ BH. In the absence of irradiation, the models produce a pronounced, observationally inconsistent gap near 10 h, wherein most systems expand, and only a minority undergo orbital decay. Including irradiation reduces or eliminates this gap, while raising the predicted accretion rates by approximately an order of magnitude. Quantitative agreement with observations is summarized in Table S2 (online). The observed BH LMXB population — which is predominantly transient — is best reproduced with a lower irradiation efficiency of $f \sim 10^{-5}$–$10^{-4}$. This range coincides with that found for transient NS LMXBs, indicating that the correlation between source type and irradiation efficiency is not specific to the accretor type but reflects a common physical origin. We note that these results represent a preliminary test. A more comprehensive validation, including a detailed comparison of the orbital period distribution through population synthesis, will be necessary to fully establish the applicability of the framework to BH LMXBs.

This framework may also be extended to cataclysmic variables (CVs). Recently, Barraza-Jorquera et al. [120] reproduced the period gap of CVs, spanning 147–191 min [121]. They adopted the boosted and disrupted MB prescription from Belloni et al. [26] and combined it with the Matt15 model. They found that simply increasing the efficiency of the Matt15 prescription by a factor of 20 reproduces the observed upper edge of the period gap. However, the physical origin of this enhancement remains unclear in their treatment. In our model, irradiation-driven winds naturally enhance MB efficiency by one to two orders of magnitude, as demonstrated in Fig. S1 (online) for efficiency factors $f = 10^{-5}$–$10^{-3}$; this offers a plausible physical explanation for the empirically required boost. Moreover, previous studies have suggested that irradiation plays a key role in modulating MT in CVs [28,122]. Therefore, our proposed framework, which couples irradiation feedback with AM loss, is likely to provide a consistent explanation for the evolution of CVs as well. A dedicated study of CVs within this unified picture, along with the population synthesis for BH systems mentioned above, will be pursued in future work.

### 4.2. The role of irradiation efficiency $f$

A central finding is that the irradiation efficiency $f$ governs evolutionary diversity within this framework. When varied over a physically plausible range ($10^{-5} \lesssim f \lesssim 10^{-3}$), $f$ maps the model predictions continuously onto the full observed spectrum of accretion rates and binary populations. This parametric simplicity, despite the underlying complexity, reveals an emergent order: the response of the donor's wind to external irradiation condenses into a single efficiency scale.

However, this efficiency scale encompasses multiple layers of physical uncertainty. By definition (Section 2.2 and Ref. [39]), $f$ is the product of three factors: the radiative efficiency of accretion $f_x$, the geometric irradiation fraction $f_\Omega$, and the net wind-driving efficiency $f_\epsilon$. Each factor is poorly constrained: $f_x$ depends on the details of the accretion flow, $f_\Omega$ is strongly affected by anisotropic X-ray emission and shadowing by the accretion disk (or a super-Eddington outflow), and $f_\epsilon$ — though stellar atmosphere calculations bracket it in the range 0.001–0.1 [37] — remains subject to considerable uncertainty. Furthermore, $f$ is not a static input parameter; it enhances the donor wind and amplifies MB, which in turn reshapes orbital evolution, the MT rate, and disk structure. Consequently, the effective $f$ can evolve dynamically with the binary state, evading a simple static prescription.

This feedback loop naturally induces MT cycles in LMXBs and CVs, a phenomenon recognized in earlier work [29,122,123]. Once MT commences, accretion-powered X-rays overshadow a fraction of the donor's intrinsic surface luminosity, driving the star out of thermal equilibrium. On short timescales, the unilluminated surface remains largely unaffected, but over the thermal timescale of the convective envelope $\tau_{ce}$, the star adjusts its radius, effective temperature, and nuclear luminosity toward a new equilibrium under the modified outer boundary condition [124,125]. During this relaxation, irradiation initially boosts the thermal expansion rate, increasing the MT rate, which further enhances accretion luminosity and irradiation, creating a positive feedback loop. However, this loop cannot operate indefinitely: after roughly a Kelvin–Helmholtz timescale $\tau_{KH}$, thermal relaxation saturates, the expansion rate drops, and the MT rate begins to decline. The feedback then reverses — irradiation weakens, the intrinsic luminosity recovers, the star shrinks, and MT decreases further. In extreme cases, MT can even cease, rendering the system detached until the next cycle begins [29].

Our framework of irradiation-boosted winds and MB adds a new dimension to this picture. When irradiation amplifies the donor wind and thereby strengthens MB, two typical regimes can emerge. If the MT rate is well below the Eddington limit, the wind-driven enhancement of braking builds up, generating a positive feedback analogous to the thermal cycle. However, because the MB boost saturates as the luminosity approaches the Eddington limit (as illustrated in Fig. 1), the feedback is naturally capped, and the system can settle into a quasi-equilibrium state without pronounced cycles. Conversely, if MT-driven orbital expansion dominates — as in sources like 2A 1822–371 — the donor's radius may expand more slowly than the orbit, causing the MT rate to drop rapidly. Subsequent MB and gravitational radiation then shrink the orbit again, reactivating MT and establishing a cycle that manifests as oscillations in the orbital period (see, e.g., Ref. [123], Figs. 2 and 6). Thus, irradiation-driven cycles can operate via both thermal relaxation and orbital AM competition, with their relative importance depending on system parameters.

This cyclic behavior has direct implications for the spin evolution of NSs. During phases of high accretion rate, the NS undergoes spin-up. However, when the MT rate drops substantially (potentially far below $10^{-12} M_\odot\,\mathrm{yr}^{-1}$, see [123] Fig. 2), the propeller effect may engage and significantly spin down the NS [126]. According to the calculations of [123] (see their Fig. 9), under cyclic MT the NS spin period should lie in the range of roughly 1 ms to several hundred milliseconds. This naturally provides a consistent explanation for the spin period of ~0.59 s observed in 2A 1822–371.

Thus, the irradiation efficiency $f$ represents far more than a simple efficiency factor; it bridges radiative input, MT, MB, stellar thermal balance, and orbital evolution. Determining its effective value as a fixed constant proves challenging from both theoretical and observational perspectives: accretion-disk contamination masks spectral signatures of the donor wind, and the inherent cyclic nature of MT implies that instantaneous measurements do not reflect the long-term average. Yet it is precisely this interplay among multiple physical processes — equilibrium and cycles, expansion and contraction, spin-up and spin-down — that generates the rich observational diversity of LMXBs and their descendants, the MSP binaries. Within this picture, the transient/persistent dichotomy, originally a classification of observational phenomenology, acquires a natural physical origin: it reflects a difference in the stability of irradiation, governed by the competition between the thermal relaxation timescale of the donor atmosphere and the temporal coherence of the irradiating flux. While other factors (such as spectral hardness or donor composition [18,127]) may also contribute, the present framework directly connects this dichotomy to irradiation stability rather than to its peak intensity, offering a unified perspective.

### 4.3. Uncertainty in the convective turnover timescale

The reference MB model (Matt15) incorporates a key parameter: the convective turnover timescale $\tau_c$. This parameter not only directly influences the braking strength (specifically in the unsaturated regime, as Eq. (8) shows that it has no effect when the field is saturated) but also determines the critical spin period $P_{crit}$, which demarcates the saturated and unsaturated regimes. However, the precise value of $\tau_c$ remains uncertain. Two primary approaches are used to calculate it: one involves integrating the path $\int_{R_b}^{R_2}(\mathrm{d}r/v_c)$ through the convective envelope — where $R_b$ is the radius at the base of the convective envelope and $v_c$ is the convective velocity (representing the travel time of a convective element from the top to the bottom of the envelope; [9,16]), while the other relies on observationally calibrated empirical relations [78,128]. In this study, we calculate $\tau_c$ using the empirical formula from Sadeghi Ardestani et al. [128] (SA17).

To assess the impact of different calculation methods, Fig. S6 (online) presents the evolutionary tracks of an example binary (with initial parameters identical to those in Fig. 1) using the above two $\tau_c$ prescriptions. The results indicate that the empirical and MESA-computed $\tau_c$ values are very similar: initially, the MESA value is slightly larger than the empirical one, though both are around 40–50 d. As the binary evolves, this relative order inverts, yet the two values remain nearly identical. The initial critical spin period, $P_{crit}$, is approximately 10 d, which is longer than the initial orbital period; thus, the magnetic activity initially resides in the saturated state. However, as the star evolves, $\tau_c$ decreases by nearly an order of magnitude. This causes $P_{crit}$ to drop below 1 d — shorter than the orbital period — so the donor's magnetic field enters the unsaturated regime, where $\tau_c$ modulates the braking strength. At this stage, the MESA $\tau_c$ remains slightly larger, resulting in marginally stronger MB, which accounts for the earlier RLO seen in the HR diagram. Because of this earlier RLO (Fig. S6c online), the donor possesses a slightly lower helium-core mass at the onset of MT, ultimately yielding a lower-mass helium WD. Owing to the inverse mass-radius relation of WDs, this leads to a larger remnant radius, a slightly higher minimum orbital period (Fig. S6a online), and a marginally lower accretion rate during the UCXB phase (Fig. S6d online).

Nevertheless, after RLO, $\tau_c$ gradually increases, while the orbital period continues to shrink. Throughout the subsequent evolution, the donor's spin period remains below $P_{crit}$, keeping the field saturated and rendering the braking strength independent of $\tau_c$. Overall, while adopting different $\tau_c$ prescriptions introduces some variations, Fig. S6 (online) demonstrates that the impact is limited, and the evolutionary paths remain largely consistent.

## 5. Conclusion

The SMB + IDW model provides a unified framework that addresses three long-standing discrepancies in LMXB evolution. It reconciles the mismatch between predicted and observed accretion rates, resolves the "fine-tuning" problem in UCXB formation, and explains the anomalous orbital expansion observed in systems such as 2A 1822–371. A single physical mechanism — the coupling of magnetic saturation with irradiation-driven winds — thus accounts for diverse phenomena across both NS and BH systems. Furthermore, when incorporated into population synthesis, the model successfully reproduces the observed orbital period distribution of binary MSPs with helium WD companions, providing independent validation that irradiation-driven winds govern both the MT phase and the final orbital architecture.

By demonstrating that irradiation feedback self-consistently modulates AM loss, the SMB + IDW framework supplants the traditional view of MB as an externally imposed condition with a description that is dynamically coupled to the accretion history. This establishes a physics-based foundation for next-generation population synthesis, predictions of gravitational-wave progenitors, and understanding the origin of MSP orbits. More broadly, this work illustrates a principle likely applicable beyond LMXBs: in close binaries, the radiation that reveals MT may also regulate it through its influence on the donor's AM budget.

## Conflict of interest

The authors declare that they have no conflict of interest.

## Acknowledgments

This work was supported by the CAS Project for Young Scientists in Basic Research (YSBR-148), the National Natural Science Foundation of China (12225304, 12041301, 12603062, 12121003, 12288102 and 12273105), the Postdoctoral Science Foundation of China (2025M783233 and GZC20261716), the National Key Research and Development Program of China (2021YFA0718500, 2021YFA1600404 and 2023YFA1607902), the Yunnan Revitalization Talent Support Program (Yunling Scholar Project), the Yunnan Science and Technology Program (202401AV070006, 202501AS070005, 202605AS350010 and 202601BC070011), and the International Centre of Supernovae (ICESUN), Yunnan Key Laboratory of Supernova Research (202505AV340004). We would like to express our gratitude to the editor and the anonymous reviewers for their time, insightful comments, and valuable guidance throughout the review process. We also thank Yudong Nie and Zonglin Yang for their helpful discussions.

## Author contributions

Zhuling Deng conceived the theoretical model, performed the numerical simulations, analyzed the data, and wrote the manuscript. Xiangdong Li and Bo Wang contributed to the conception of the theoretical model, participated in the analysis of the results, assisted in revising the manuscript, and supervised the project. Philipp Podsiadlowski and Zhanwen Han analyzed the results, contributed to the interpretation and provided important suggestions. Yong Shao and Dongdong Liu assisted in revising the manuscript. All authors discussed the results and approved the final version of the manuscript.

## Appendix A. Supplementary material

Supplementary data to this article can be found online at https://doi.org/10.1016/j.scib.2026.08.068.